\documentclass{elsart}
\usepackage{amssymb}
\usepackage{amsmath}
\usepackage[dvips]{graphicx}

\journal{\textbf{Colloids \& Surfaces A}}
\input{tcilatex}
\begin{document}

\begin{frontmatter}

\title{ {\large  Liquid-solid interaction at nanoscale \\ and its application in vegetal biology}}

\author{Henri Gouin}
\ead{henri.gouin@univ-cezanne.fr,  \emph{Telephone:}  +33 491 288 407, \emph{Fax:}  +33 491 288 776}

\address { C.N.R.S. U.M.R. 6181 \&
 University of Aix-Marseille, \\ Case 322, Av. Escadrille
 Normandie-Niemen, 13397 Marseille Cedex 20 France}

\begin{abstract}+
The water ascent in tall trees is   subject to controversy: the vegetal biologists  debate on the validity of the \emph{cohesion-tension theory}  which considers strong negative pressures in microtubes of xylem carrying the crude sap.
This article aims to point out that liquids are submitted at the walls to intermolecular forces inferring density gradients  making heterogeneous liquid layers  and therefore  disqualifying the Navier-Stokes equations for nanofilms. The crude sap motion takes  the disjoining pressure gradient into account  and the sap flow  dramatically increases   such that the watering of nanolayers may be analogous to a microscopic flow.
Application to microtubes of xylem avoids the problem of cavitation  and enables us to understand why the ascent of sap is possible for very high trees.

\end{abstract}


\end{frontmatter}

\section{Introduction}

The model we develop makes it possible to investigate the behavior of the
fluids in the nanofilms, and its applications extend to life sciences. A
particularly interesting example concerns vegetable biology: the rise and
the motion of  sap in the highest trees focus many polemics and debates
between biologists. Many of them regard the approach known as
the \emph{cohesion-tension theory (CTT)} proposed at the end of the
nineteenth century by Dixon and Joly  as the only valid one \cite{Dixon}.

As an obvious fact, Flindt reports huge trees as eucalyptus and  giant
sequoias of more than 130 meters \cite{Flindt}, but the  biophysical
determination of maximum size to which trees can grow is not well
understood and calculated. The main problem with the understanding of water
transport is why the sap is able to irrigate up very high
levels in tall trees. \newline
The crude sap contains diluted salts  but its physical properties are
roughly comparable with the water ones. Hydrodynamics, capillarity and
osmotic pressure  create a ascent of sap of only few tens of meters \cite{Zimm}%
. To explain the sap ascent phenomenon, Dixon and Joly proposed a
cohesion-tension model, followed by a quantitative attempt \cite{Honert}:
liquids are assumed to be subjected to tensions generating negative
pressures compensating gravity effects. \newline
As pointed out in \cite{Zimm2}, a turning-point in the confidence of the
opponents to the cohension-tension theory for the sap ascent was the
experiment which demonstrated that tall trees survive by overlapping double
saw-cuts made through the cross-sectional area of the trunk to sever all
xylem elements \cite{Preston}. This result confirmed by several authors does
not seem in agreement with the possibility of strong negative pressures in
microtubes \cite{Mackay,Benkert}. Using a xylem pressure probe, the
apparatus does not measure any water tension in many circumstances: xylem
tension exceeding 0.6 Mpa seems not to be observed and in normal state most
vessels may be embolized at a level corresponding about sixty meter height.
Moreover, gas-vapor transportation in xylem tubes seems to appear at the top
of high trees (\cite{Zimm2} and its references herein).\newline
As comments and questions,  M.H. Zimmerman wrote in 1983 \cite{Zimm}: \newline
"We don't yet fully understand all aspects of xylem-water supply to leaves
and have here a wide-open filed of potential very interesting future
research.
The heartwood is referred to as a wet wood. It may contain liquid under
positive pressure while in the sapwood the transpiration stream moves along
a gradient of negative pressures. Why is the water of the central wet core
not drawn into the sapwood?
Free water, i.e. water in tracheids, decreases in successively older layers
of wood as the number of embolized tracheids increases. The heartwood is
relatively dry i.e. most tracheids are embolized.  It is rather ironic that
a wound in the wet wood area, which bleeds liquid for a long period of time,
thus appears to have the transpiration stream as a source of water, in spite
of the fact that the pressure of the transpiration stream is negative most
of the time!
It should be quite clear by now that a drop in xylem pressure below a
critical level causes cavitations and normally puts the xylem out of
function permanently. The cause of such a pressure drop can be either a
failing of water to the xylem by the roots, or excessive demand by
transpiration." \newline
Many proponents of the \emph{CTT} wrote a letter \cite{all} to protest
against the recent  review \cite{Zimm2}. They said that "the \emph{CTT} is
widely supported by biological scientists as the only theory consistent with
the preponderance of data on water transport of plant". \newline
Nonetheless, the problem of possible cavitation in trees remains. Such
liquids are strongly metastable and can generate cavitations causing
embolisms in xylem tubes made of dead cells \cite{Tyree}. For example, it is
interesting to note that in xylem tube - where diameters range between 50
and 400 $\mu m$ - the crude sap has a surface tension $\gamma_{lv}$
lower than the surface tension of pure water which is $72$ cgs at 20${%
{}^\circ}$ Celsius. If we consider a microscopic gas-vapor bubble inside the
crude sap with  diameter $2\,R$, the
difference between the gas-vapor pressure $P_{vapor}$ and the liquid sap pressure $%
P_{liquid}$ can be expressed by the Laplace formula: $\displaystyle P_{vapor}-P_{liquid} =
2\,\gamma_{lv}/R$. But $P_{vapor}$ being positive, unstable bubbles must appear
when $\displaystyle
R\geq - 2\,\gamma_{lv}/P_{liquid}$. For a negative pressure $P_{liquid} \leq -0.6$ MPa
corresponding to more than sixty meters height, we get $R \geq 0.24\, \mu m$. In such a
case, dynamical bubbles spontaneously appear from germs naturally existing
in a crude liquid and  cavitation makes the tubes embolized. Consequently,
without any biological  known process it is difficult to be convinced that
xylem tubes  are not embolized when they are filled enough with sap up to
altitude significantly more important  than one hundred meters corresponding
to the highest trees.

Our understanding of the ascent and the motion of sap in very high trees
differs from the  \emph{CTT}:  at  a higher level than a few  tens of
meters - corresponding to the pulling of water by capillary and osmotic
pressure -  we assume that xylem microtubes are  embolized. In addition, we
also assume that a thin liquid film - with a thickness of a few  nanometers
\cite{Derjaguin,Israel} - wets xylem  walls up to the top of the tree. At
this scale, long range molecular forces stratify liquids and  the ratio
between tube diameter and  sap film thickness allows us to consider tube
walls  as plane surfaces. \newline
In Section 2, using the calculations presented in \cite{Gouin3,Gouin2}, we
reconsider the  analytic expression in density-functional theory for a thin
heterogeneous liquid film which takes    account of the power-law tail behavior   dominant in a thin liquid film
in contact with a solid \cite{gouin}. The effects of the vapor bulk
bordering the liquid film  are simply expressed with an other
density-functional located on a mathematical surface. With such a
functional, we obtained the equations of equilibrium, motion and boundary
conditions \cite{Gouin1} for a thin vertical liquid film wetting a vertical
solid wall and we computed the liquid layer thickness as a function of the
film level; these previous results can be extended to mixtures of fluid and perfect gas \cite{Gouin3}%
. Then, the so-called \emph{disjoining pressure} of thin liquid layers
yields a natural tool for very thin films \cite{Derjaguin}. The minimal
thickness for which a stable wetting film wets a solid wall is associated
with the \emph{pancake layer} when the film is bordering the dry solid wall
and corresponds to the maximal altitude \cite{Derjaguin,de Gennes1,de
Gennes2}. The normal stress vector acting on the wall remains constant through
the liquid layer and corresponds to the gas-vapor bulk pressure which is
currently the atmospheric pressure and consequently, no negative pressure
appears in the liquid layer. At the top of very high trees, the thickness of
the sap layer is of a few number of nanometers. The negative pressure is
only present for the liquid bulk in micropores. Numerical calculations
associated with physical values for water yield the maximal film altitude
for a wood material corresponding to a good order of the height of the
tallest trees.  \newline
In Section 3, we consider the flow of sap at high levels. For shallow water,
the flows of liquids on solids are mainly represented by using the
Navier-Stokes equations associated with adherence conditions at the walls.
Recent experiments in nanofluidics seem to prove, also for liquids, that at
nanoscales corresponding to sap layers at very high tree levels, the
conditions of adherence are disqualified \cite{degennes,Tabeling}. With the
aim of explaining experimental results, we reconsider the fluids as media
whose motions generate slips along the walls; so, we can draw consequences
differing from results of classically adopted models as reconsidered in \cite%
{gouin7}. The new model we are presenting reveals an essential difference
between the flows of microfluidics and those of nanofluidics. In the latter,
simple laws of scales cannot be only taken anymore into account.\newline
The transpiration in the leaves induces a variation of the sap layer
thickness in microtubes. Consequently, the gradient of thickness along
microtubes creates a gradient of disjoining pressure which induces driving
forces along the layer. For thin layers, the sap flow depending on the
variations of the layer thickness can be adapted to each level of leaves
following the tree requirement. This is an important understanding why the
flow of sap can be non negligible at a level corresponding to the top of the
tallest trees. Moreover, we notice that the stability criterium of the flow issued
from the equation of motion fits with the results of Derjaguin's school \cite%
{Derjaguin}.

\section{A study of inhomogeneous fluids near a solid wall}

In this section, we recall the main results presented in \cite{Gouin3,Gouin2}%
. Thanks to these results, in Section 3, we shall consider sap layers of the
highest trees with a thickness of some nanometers only.\newline
The density-functional of an inhomogeneous fluid in a very thin isothermal
layer domain $(O)$ of wall boundary $(S)$ and liquid-vapor interface $(\Sigma)$
was chosen in the form:
\begin{equation}
F = \int\int\int_{(O)} \varepsilon\ dv + \int\int_{(S)} \phi\ ds +
\int\int_{(\Sigma)} \psi\ ds.  \label{density functional2}
\end{equation}

$\bullet$ The first integral is associated with a square-gradient
approximation when we introduce a specific free energy of the fluid at a
given temperature $T$ as a function of density $\rho$ and $\beta= (\mathrm{%
grad\, \rho)^2}$  such as \cite{Widom,rowlinson}:
\begin{equation*}
\rho \,\varepsilon =\rho \,\alpha (\rho)+\frac{\lambda }{2}\,(\text{grad\ }%
\rho )^{2},  \label{internal energy}
\end{equation*}
where term $({\lambda }/{2})\,(\mathrm{grad\ \rho )^{2}}$ is added to the
volume free energy $\rho \,\alpha (\rho)$ of a compressible fluid and scalar $%
\lambda$ is assumed to be constant at a given temperature \cite{Rocard}.
Specific free energy $\alpha $ enables liquid and vapor bulks to be
continuously connected and the  pressure $P(\rho)=\rho ^{2}\alpha _{\rho
}^{\prime }(\rho )$ is similar to van der Waals one.

$\bullet$  For a plane solid wall $(S)$, the solid-liquid  surface free
energy is in the form \cite{gouin,Weiss}:
\begin{equation}
\phi(\rho)=-\gamma _{1}\rho+\frac{1}{2}\,\gamma_{2}\,\rho^{2}.
\label{surface energy}
\end{equation}
Here $\rho$ denotes the fluid density value at surface $(S)$; constants $%
\gamma _{1}$, $\gamma _{2}$ and $\lambda$ are generally positive and given
by the mean field approximation:
\begin{equation}
\gamma _{1}=\frac{\pi c_{ls}}{12\delta ^{2}m_{l}m_{s}}\;\rho _{sol},\quad
\gamma _{2}=\frac{\pi c_{ll}}{12\delta^2 m_{l}^{2}},\quad \lambda = \frac{%
2\pi c_{ll}}{3\sigma_l \,m_{l}^{2}},  \label{coefficients}
\end{equation}
where $c_{ll}$ and $c_{ls}$ are two positive constants associated with
Hamaker constants; $\sigma_l$   and $\sigma_s$ denoting fluid and
solid molecular diameters, $\delta = \frac{1}{2}(\sigma_l+\sigma_s)$; $%
m_{l}$, $m_{s}$ denote masses of fluid and solid molecules; $%
\rho _{sol}$ is the solid density.

$\bullet$ For the plane liquid-vapor interface $(\Sigma)$ the surface
free energy $\psi$ is reduced to \cite{Gouin3,Gouin2}:
\begin{equation*}
\psi (\rho )=\frac{\gamma _{4}}{2}\ \rho ^{2},  \label{cl2}
\end{equation*}
where  $\rho$ is the density of the liquid bounding the interface and $\gamma_4$ is associated
with the interfacial thickness of the order of the fluid molecular
diameter ($\gamma_4\simeq \gamma_2 $).

In case of equilibrium, functional (\ref{density functional2}) is stationary
and yields the \emph{equation of equilibrium} and \emph{boundary
conditions} \cite{Gouin1,Pismen}.

\subsection{Equation of equilibrium}

The equation of equilibrium is \cite{Gouin1,gouin4}:
\begin{equation}
\text{div }\mathbf{\sigma} + \rho\, g\, \mathbf{i}=0\,,
\label{equilibrium1a}
\end{equation}
where \,$ \mathbf{\sigma =}-\left(\rho
^{2}\varepsilon _{\rho }^{\prime }-\rho \text{ div\textrm{\ }}(\lambda \text{
grad }\rho )\right)\mathbf{1}-\lambda \;\text{grad\
}\rho \ \otimes \ \text{grad }\rho,$ $g$ is the acceleration of gravity and $\mathbf{i}$, of coordinate  $x$, is the downward direction. Let us
consider an isothermal vertical film of liquid; then in orthogonal system,
the coordinate $z$ being external and normal to the flat vertical solid wall,
spatial density derivatives \emph{are negligible} except in direction of $z$%
. In the complete liquid-vapor layer (we call \emph{interlayer})  and along
direction $z$, Eq. (\ref{equilibrium1a}) yields a constant value at each
level $x$:
\begin{equation*}
P(\rho)+\frac{\lambda }{2}\left(\frac{d\rho }{dz}\right)^2-\lambda\, \rho\, \frac{%
d^2\rho}{dz^2}=P_{v_{b_x}},  \label{equilibrium1b}
\end{equation*}
where $P_{v_{b_x}}=P(\rho_{v_{b_x}})$ denotes the pressure in the vapor
bulk of density $\rho_{v_{b_x}}$ bounding the liquid layer at level $x$. In
the fluid, Eq. (\ref{equilibrium1a}) can  be written \cite{gouin4}:
\begin{equation}
\text{grad}\left( \mu_o -\lambda\, \Delta \rho - g\,x \right) =0 ,
\label{equilibrium2a}
\end{equation}
where $\mu_o $ is the chemical potential (at temperature $T$), chosen null in     the
liquid and vapor bulks of phase
equilibrium densities $\rho _{l}$ and $\rho _{v}$, respectively; $\Delta$ denotes the Laplacian.  Thanks to Eq. (\ref{equilibrium2a}), we obtain in all the
fluid \emph{and not only in the interlayer}:
\begin{equation}
\mu _{o}(\rho)-\lambda \Delta \rho - g\ x =\mu _{{o}}(\rho _{b}) ,
\label{equilibrium2b}
\end{equation}
where $\mu _{{o}}(\rho _{b})$ is the chemical potential value of a \emph{%
liquid mother bulk} of density $\rho _{b}$  such that $\mu _{{o}}(\rho
_{b})= \mu _{{o}}(\rho_{v_{b}})$, where $\rho_{v_{b}}$ is the density of the
\emph{vapor bulk} bounding the layer \emph{at level} $x =0$ \cite%
{Derjaguin}. Equation (\ref{equilibrium2b}) is valid in the  interlayer
and yields the equation of density profile:
\begin{equation}
\lambda\,\frac{d^2\rho}{dz^2} = \mu_{b_x}(\rho), \quad \mathrm{with}\quad
\mu_{b_x}(\rho) = \mu_o(\rho)-\mu_o(\rho _{b_x}) .  \label{equilibrium2d}
\end{equation}
where $\rho _{b_x}$ is the liquid mother bulk density at level $x$.

\subsection{The disjoining pressure for vertical liquid films}

The disjoining pressure at level $x$ can be written as \cite{Derjaguin}:
\begin{equation*}
\Pi =P_{v_{b_x}}-P_{b_x}\,,  \label{disjoining pressure}
\end{equation*}
where $P_{b_x}= P(\rho_{b_x})$. At a given
temperature $T$, $\Pi$ is a function of $\rho_{b_x}$. The reference chemical
potential linearized near $\rho_l\,$ is $\ \mu _{o}(\rho)= ({c_{l}^{2}}/{\rho
_{l}})(\rho -\rho_{l})\ $ where $c_l\,$ is the isothermal sound velocity in
liquid bulk $\rho_l$ at temperature $T$ \cite{espanet}. In the liquid part
of the liquid-vapor film,  Eq. (\ref{equilibrium2d}) writes:
\begin{equation}
\lambda \frac{d^{2}\rho }{dz^{2}} =\frac{c_{l}^{2}}{\rho _{l}}(\rho -\rho
_{b})-g\,x \equiv \frac{c_{l}^{2}}{\rho _{l}}(\rho -\rho _{b_x})\quad
\mathrm{with}\ \ \rho _{{b_x}} = \rho _{{b}}+ \frac{\rho _{l}}{c_l^2}\, g\,
x .  \label{liquidensity}
\end{equation}

At level $x=0$, the liquid mother bulk density is closely equal to $\rho_l$
and because of Rel. (\ref{liquidensity}), $\Pi$ can be considered as a
function of $x$ \cite{Gouin3}:
\begin{equation}
\Pi (x)=-\rho_l\,g\,x\left(1+\frac{g\,x}{2\,c_l^2}\right).
\label{disjoining pressuregravity}
\end{equation}
Now, we consider a film of thickness $h_x$ at level $x$; the density profile
in the liquid part of the liquid-vapor film is solution of the system:
\begin{equation*}
\left\{
\begin{array}{c}
\displaystyle\lambda \frac{d^{2}\rho }{dz^{2}}=\frac{c_{l}^{2}}{\rho _{l}}
(\rho -\rho _{b_x}), \\
\quad \mathrm{with}\quad \displaystyle\lambda \frac{d\rho }{dz}_{\left|
_{z=0}\right. }=-\gamma _{1}+\gamma _{2\ }\rho _{\left| _{z=0}\right. }\quad
\mathrm{and}\quad \displaystyle\lambda \frac{d\rho }{dz}_{\left|
_{z=h_x}\right. }=-\gamma _{4}\ \rho _{\left| _{z=h_x}\right. }.%
\end{array}
\right.  \label{systeme1}
\end{equation*}
Let quantities $\tau $, $d $ and $\gamma _{3}$ be defined as:
\begin{equation}
\tau \equiv \frac{1}{d}=\frac{c_{l}}{\sqrt{\lambda \rho _{l}}}\qquad \mathrm{%
and}\qquad \gamma _{3}\equiv\lambda \tau ,  \label{tau}
\end{equation}
such that $d $ is a reference length. Due to the fact that $\rho
_{b_x}\simeq \rho _{b}\simeq \rho _{l}$ \cite{Derjaguin}, the disjoining
pressure reduces to \cite{Gouin2}:
\begin{eqnarray}
\Pi (h_x) &=&\frac{2\,c_{l}^{2}}{\rho _{l}}\left[ (\gamma _{1}-\gamma
_{2}\rho _{l})(\gamma _{3}+\gamma _{4})e^{h_x\tau }+(\gamma _{2}-\gamma
_{3})\gamma _{4}\rho _{l}\right] \times  \notag \\
&&\frac{\left[ (\gamma _{2}+\gamma _{3})\gamma _{4}\rho _{l}-(\gamma
_{1}-\gamma _{2}\rho _{l})(\gamma _{3}-\gamma _{4})e^{-h_x\tau }\right] }{%
\left[ (\gamma _{2}+\gamma _{3})(\gamma _{3}+\gamma _{4})e^{h_x\tau
}+(\gamma_{3}-\gamma _{4})(\gamma _{2}-\gamma _{3})e^{-h_x\tau }\right] ^{2}}
.  \label{Derjaguine bis}
\end{eqnarray}
The disjoining pressure of the mixture of liquid and perfect gas is the same
than for a single van der Waals fluid and calculations and results are
identical to those previously obtained \cite{Gouin3}.

\subsection{Water wetting a vertical plane wall of xylem}

Our aim is to point out an example such that previous results provide a
value of maximum height for a vertical water film wetting a plane wall of
xylem and to estimate the sap layer thickness at this altitude. \newline
As proved by Derjaguin \emph{et al} in \cite{Derjaguin}, ({Chapter 2}),
the Gibbs free energy per unit area $G$ can be expressed as a function of  $%
h_x$:
\begin{equation*}
G (h_x) = \int_{h_x}^{+\infty} \Pi(h)\,dh,  \label{Gibbs1}
\end{equation*}
where $h_x=0$ is associated with the dry wall in contact with the vapor bulk
and $h_x=+\infty$ is associated with a wall in  contact with the liquid bulk.
The spreading coefficient is $S = \gamma_{_{SV}} -
\gamma_{_{SL}}-\gamma_{_{LV}},\label{wetting} $ where $\gamma_{_{SV}},
\gamma_{_{SL}}, \gamma_{_{LV}}$ are the solid-vapor, solid-liquid and
liquid-vapor free energies per unit area of interfaces, respectively.  The
energy of the liquid layer per unit area can be written \ $E =
\gamma_{_{SL}}+\gamma_{_{LV}}+ G(h_x). \label{layer energy} $ \newline
The coexistence of two film segments with different thicknesses is a
phenomenon interpreted with  the equality of chemical potentials and the
equality surface tensions of the two films. A spectacular case
corresponds to the coexistence of a liquid film of thickness $h_p$ and a dry
solid wall  associated with $h_x=0$. The film is the so-called \emph{pancake
layer} corresponding to the  condition \cite{Derjaguin,de Gennes1}:
\begin{equation*}
G(0) = G(h_p)+ h_p\, \Pi(h_p).  \label{pancake thickness}
\end{equation*}
Liquid films of thickness $h_x > h_p$ are stable and liquid films of
thickness $h_x < h_p$ are metastable or unstable. For a few nanometer range,
the film thickness is not exactly $h_x$; we must add the thickness estimated   at $2\,\sigma_l$ of the liquid
part of the liquid-vapor interface bordering the liquid layer  and the
film thickness is $e_x \approx h_x+ 2\, \sigma_l$ \cite{rowlinson,Rocard}. \newline
When $h_x = 0$ (corresponding to the dry wall), the value of $G$ is the
spreading coefficient $S$. Point  $P$ associated with the pancake layer is
observed on the curve to be closely an inflexion point of graph $\Pi(h_x)$
\cite{Gouin3}. To obtain the pancake thickness corresponding to the smallest
film thickness, we draw the graphs of $\Pi(h_x)$ and $G(h_x)$  when $h_x \in
[({1}/{2})\,\sigma_l,\ell]$, where $\ell$ is a distance of few tens of
Amgstr\" om.

\textbf{For the numerical calculations}, we considered water at $T= 20 {%
{}^\circ}$ Celsius wetting a wall in xylem. The experimental estimates of
coefficients are obtained in \textbf{c.g.s. units} \cite{Israel,Gouin2,Handbook}:
\newline
$\rho_l =0.998$, $c_l = 1.478\times 10^{5}$, $c_{ll}=1.4\times 10^{-58}$, $%
\sigma _{l}=2.8\times 10^{-8}\text{\ (2.8 {\aa}ngstr\"{o}ms), }$ $%
m_{l}=2.99\times 10^{-23}$.  From Rel. (\ref{coefficients}), we deduce  $%
\lambda =1.17\times 10^{-5}$,  $\gamma _2=\gamma _4 = 54.2$.  From Rel. (\ref%
{tau}), we get $\gamma _3 = 506$,  $d = 2.31 \times 10^{-8}$. \newline
We consider a material such that the Young angle between the liquid-vapor
interface and the solid material surface is $\theta \approx 50{{}^\circ}$.
This Young angle is an arithmetic average of   Young angles for water wetting different xylem
walls \cite{Mattia}. The coefficient $c_{ls}$ is obviously not given in the physical tables and $\gamma_1$ cannot be obtain from Rel. (\ref{coefficients}); due to Rel. (\ref{surface energy}), we immediately get
the   unknown coefficient $\gamma_1$  in  Rel. (\ref{Derjaguine bis}): $\gamma_1 \approx 75$.

\textbf{In Fig. 1 - left graph}, we present the disjoining pressure graph $%
\Pi(h_x)$. The physical part of disjoining pressure graph corresponding to $%
\partial\Pi/\partial h_x <0$ is a plain line  and is associated
with thickness liquid layer of several molecules. The dashed line has no real existence.  \newline
\textbf{In Fig. 1 - right graph}, we present the free energy graph $G(h_x)$.
Due to $h_x>({1}/{2})\,\sigma_l$, it is not possible to obtain the limit
point $W$ corresponding to the dry wall. This point is obtained by an
interpolation associated with the concave part of the \emph{G}-curve. Point $%
P$ follows from the drawing of the tangent line issued from $W$  to the \emph{G}-curve. The limit of the film thickness is
associated to the \emph{pancake} thickness  $e_p\approx h_p + 2\,\sigma_l$
when the liquid film coexists with the dry wall.  The reference length $d$
is of the same order than $\sigma_l,\, \sigma_s$ and $\delta$ and is a good
length order for very thin films. The total pancake thickness $e_p$ is of
one nanometer order corresponding to a good thickness value for a
high-energy surface \cite{de Gennes2}. We deduce $S \approx 40$ \texttt{cgs}.  However, crude sap is not pure water. Its liquid-vapor surface tension
has a lower value than surface tension of pure water (72 \texttt{cgs} at $20{%
{}^{\circ }}$C) and it is possible to obtain the same spreading coefficients
with less energetic surfaces. \newline
When $|x|$ is of some hundred meters, Eq. (\ref{disjoining pressuregravity})
yields \ $\Pi(x)\simeq -\rho_l\,g\,x . $ \newline
The maximum of altitude $|x_{_M}|$ corresponds to the pancake layer. To this
altitude, we add 20 meters corresponding to the ascent of sap due to capillarity and osmotic pressure and we obtain a film height
of $140$ meters which is of the same order than the topmost trees. We also
note the important result: in the trees, the thickness of the layer is of some
nanometers at high level.
\begin{figure}[h]
%\begin{center}
\includegraphics[width=14.5cm]{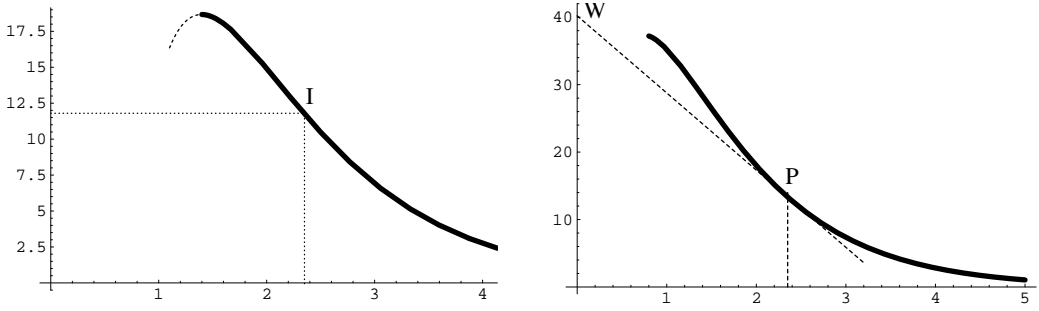} %\end{center}
\caption{\emph{\textbf{Left graph}: $\Pi(h_x)$-graph. The unit of $x-$axis
is $d=2.31\times 10^{-8}$ \texttt{cm}; the unit of $y-$axis is one
atmosphere. \textbf{Right graph}: $G(h_x)$-graph. The unit of $x-$axis is $%
d=2.31\times 10^{-8}$ \texttt{cm} ; the unit of $y-$axis is one \texttt{cgs}
unit of surface tension. }}
\label{fig1}
\end{figure}

\section{Dynamics of liquid nanofilms}

The idea of xylem tubes completely filled with sap induces that the flow of
liquid  along vessels can be compared with a flow through capillaries \cite%
{Zimm}. The flow rate through capillary tubes is proportional to the applied
pressure gradient and the hydraulic conductivity. Parabolic flow causes the
flow rate to be proportional to the fourth power of capillary radius \cite%
{batchelor}. One of the most important design requirements is that vapor
blockage does not happen in the stem. Consequently, to be efficient for the
transport of sap, the xylem tube radius must be as larger as possible, which is not the case.
\newline
The dynamics of such liquid nanofilms is always studied in isothermal case.
Our model of sap nanolayer implies different consequences:
 \\
-\ The classical model by Navier-Stokes is unable to describe fluid
motions in very thin films,   \\
-\ The notion of surface free energy of a sharp
interface separating gas and liquid layer must be reconsidered,\newline
-\ The equation of fluid motions along the nanofilm is obtained by adding the
forces of viscosity to the conservative forces, \newline
-\ The evolution equation of the film thickness   takes account of the variation of the disjoining pressure along the layer.\newline
At equilibrium, the different fluid quantities are $\rho
_{b_{x}}, \rho_{v{
_{b_{x}}}}, P_{b_{x}}, P_{v{
_{b_{x}}}}, h_{x} \ldots $ For the motion, the corresponding quantities are
denoted by $\rho^{\ast } _{b_{x}},\rho^{\ast }_{v{
_{b_{x}}}}, P^{\ast }_{b_{x}}, P^{\ast }_{v{
_{b_{x}}}}, h_{x}^{\ast } \ldots $\newline
When the liquid layer thickness is small with respect to transverse
dimensions of the wall, it is possible to simplify the Navier-Stokes
equation which governs the flow of  classical viscous fluids. \\
When $h\ll L$,
where $L$ is the wall transversal characteristic size,\\ \emph{i)} The
velocity component along the wall is large with respect to the normal
velocity component which can be neglected,\\
\emph{ii)} The velocity vector mainly
varies  along the direction orthogonal to the wall and it is possible
to neglect velocity derivatives with respect to coordinates along the wall
compared to the normal derivative,\\
\emph{iii)} The pressure is constant in
the direction normal to the wall. It is possible to neglect the inertial
term when $Re\ll L/h$, where $Re$ is the Reynolds number of the flow.\newline
Equation of hydrodynamics is not valid in a liquid nanolayer because the
fluid is heterogeneous and the liquid stress tensor is not anymore scalar.
However, it is possible to adapt the results obtained for viscous flows to motions in liquid nanolayers: due to $\ \epsilon =h/L\ll 1$, we are also in the case of \emph{%
long wave approximation}. \newline
We denote the velocity by $\mathbf{V}=(u,v,w)$ where $(u,v)$ are the
tangential components. Due to the fact that $\,e=\mathrm{sup}\left(
\left\vert w/u\right\vert ,\left\vert w/v\right\vert \right) \ll 1$, we also are
 in the case of \emph{approximation of lubrication}. The main parts of
terms associated with second derivatives of liquid velocity components
correspond to ${\partial ^{2}u}/{\partial z^{2}}$ and ${\partial ^{2}v}/{%
\partial z^{2}}$. The density is constant along each stream line ($\overset{%
\mathbf{\centerdot }}{\rho }=0\Longleftrightarrow div\,\mathbf{V}=0$) and
isodensity surfaces contain the trajectories. Then, ${\partial u}/{\partial
x},{\partial v}/{\partial y}$ and ${\partial w}/{\partial z}$ have the same
order of magnitude and $\epsilon \sim e$.\newline
As in \cite{Rocard}, we assume that the kinematic viscosity coefficient $\nu
=\kappa/\rho $ only depends  on the temperature. In motion equation, the
viscosity term is
$$\ ({1}/{\rho })\,\text{div } \mathbf{\sigma }_{v} = 2\nu \,%
\left[ \ \text{div }\textbf{\emph{D}}\,+\,\textbf{\emph{D}}\text{ grad \{\thinspace Ln}\,(2\,\kappa)\}\ %
\right],  $$
where $\emph{\textbf{D}}$ is the velocity  deformation tensor and $\emph{\textbf{D}}$ grad\{Ln ($2\,\kappa$)\} is negligible with respect to
div\thinspace $\textbf{\emph{D}}$. \newline
In both lubrication and long wave approximations, the liquid nanolayer
motion verifies \cite{gouin4}:
\begin{equation*}
{\mathbf{\Gamma }}+\text{grad}[\,\mu _{o}(\rho )-\lambda \,\Delta \rho
\,]=\nu \,\Delta {\mathbf{V}}+g\,{\mathbf{i}}\ \ \mathrm{with}\ \ \Delta {%
\mathbf{V}}\simeq
\begin{bmatrix}
\displaystyle\;\frac{\partial ^{2}u}{\partial z^{2}},\displaystyle\;\frac{%
\partial ^{2}v}{\partial z^{2}},0%
\end{bmatrix}%
.  \label{motion0}
\end{equation*}%
This equation corresponds to the equation of equilibrium ({\ref%
{equilibrium1a}) with addition of inertial (acceleration) term $\mathbf\Gamma$ and viscous term $\nu \,\Delta {\mathbf{V}}$.\newline
In approximation of lubrication, the inertial term is neglected:
\begin{equation}
\text{grad}[\,\mu _{o}(\rho )-\lambda \,\Delta \rho
\,]=\nu \,\Delta {\mathbf{V}}+g\,{\mathbf{i}}.  \label{motion}
\end{equation}%
Equation ({%
\ref{motion}) separates into tangential and normal components to the solid
wall.

-\ The normal component - following $z$ - of Eq. (\ref%
{motion}) writes in the same form than for equilibrium:
\begin{equation*}
\frac{\partial }{\partial z}\left[ \ \mu _{o}(\rho )-\lambda \,\Delta \rho \,%
\right] =0,
\end{equation*}%
and consequently,
\begin{equation}
\mu _{o}(\rho )-\lambda \,\Delta \rho =\mu _{o}(\rho^{\ast } _{b_{x}}),
\label{motion h}
\end{equation}%
where $\rho^{\ast } _{b_{x}}$ is the dynamical  liquid mother bulk density at level $x$
(different from the liquid bulk density $\rho _{l}$ of the plane interface
at equilibrium and also from $\rho _{b_{x}}$, liquid mother bulk density at level $x
$ at equilibrium).  \\
To each density $\rho^{\ast } _{b_{x}}$ is associated a liquid nanolayer
thickness $h_{x}^{\ast }$. We can write $\mu _{o}(\rho _{b})-\lambda
\,\Delta \rho =\eta (h_{x}^{\ast }),$ where $\eta $ is such that $\eta
(h_{x}^{\ast })=\mu _{o}(\rho^{\ast}_{b_{x}})$.

-\ For one-dimensional motions colinear to the solid wall (direction ${\mathbf{i%
}}$ and velocity $u\,{\mathbf{i}}$), by taking  account of Eq. (\ref%
{motion h}), the tangential component of Eq. (\ref{motion}) yields:
\begin{equation*}
{\mathbf{i}}\,.\,\text{grad{\ }}\mu _{o}(\rho^{\ast } _{b_{x}})=\nu \frac{%
\partial ^{2}u}{\partial z^{2}}+g\,,
\end{equation*}%
which is equivalent to:
\begin{equation}
\frac{\partial \mu _{o}(\rho^{\ast } _{b_{x}})}{\partial \rho^{\ast
} _{b_{x}}}\ \frac{\partial \rho^{\ast } _{b_{x}}}{\partial x}=\ \nu \frac{\partial
^{2}u}{\partial z^{2}}+g.  \label{viscosity}
\end{equation}%
For the most practical situations, simple fluids slip on a solid wall only at a
molecular level \cite{Churaev} and consequently, the kinematic condition at
solid walls is the adherence condition $({z=0\;\Rightarrow \;u=0})$.
Nevertheless, with water flowing on thin hydrophobic nanolayers, there are
some qualitative observations for slippage. With water flowing in thin,
hydrophobic capillaries, there are also some early qualitative evidences for
slippage \cite{Blake}. De Gennes said: "the results are unexpected and
stimulating and led us to think about unusual processes which could take
place near a wall. They are connected with the thickness $h$ of the film
when $h$ of an order of the mean free path" \cite{degennes}.\\ Recent papers  in nonequilibrium molecular dynamics simulations of   three
dimensional micro-Poiseuille flows in Knudsen regime
reconsider   microchannels:  the influence of gravity force, surface roughness,
surface wetting condition and wall density are investigated. The results
point out that the no-slip condition can be observed for Knudsen flow when
the surface is rough. The roughness is a dominant parameter as far
as the slip of fluid is concerned. The surface wetting condition  substantially influences
the velocity profiles \cite{Tabeling}.\newline
In fluid/wall slippage, the condition at solid wall writes:
\begin{equation*}
  u=L_{s}\frac{\partial u}{\partial z}  \qquad {\rm at}\  z=0,\
\end{equation*}%
where $L_{s}$ is the so-called \emph{Navier length}. The Navier length is expected to
be independent of $h$ and may be as large as a few microns \cite{Tabeling}.%
\newline
At the liquid-vapor interface, we also assume that vapor viscosity stress is
negligible; from the continuity of fluid tangential stress through a
liquid-vapor interface, we get:
\begin{equation*}
 \frac{\partial u}{\partial z}=0 \qquad {\rm at}\ z=h_{x}^{\ast }\;.
\end{equation*}%
Consequently, Eq. (\ref{viscosity}) implies:
\begin{equation*}
\nu \,u=\left( \frac{\partial \mu _{o}(\rho^{\ast } _{b_{x}})}{\partial \rho^{\ast }
_{b_{x}}}\ \frac{\partial \rho^{\ast } _{b_{x}}}{\partial x}%
-g\right) \left( \frac{1}{2}\,z^{2}-h_{x}^{\ast }\,z-L_{s}h_{x}^{\ast
}\right) .
\end{equation*}%
The mean spatial velocity $\overline{u}$ of the liquid in the nanolayer at
level $x$ is:
\begin{equation*}
{\displaystyle\overline{u}=\frac{1}{h_{x}^{\ast }}\int_{o}^{h_{x}^{\ast }}u\
dz}.
\end{equation*}%
Previous computations yield:
\begin{equation*}
\nu \,{\mathbf{\overline{u}}}=-h_{x}^{\ast }\left( \frac{h_{x}^{\ast }}{3}%
+L_{s}\right) \ \left[ \,\text{grad}\ \mu _{o}(\rho^{\ast } _{b_{x}})-g\,\textbf{i}%
\right] \qquad \mathrm{with}\qquad {\mathbf{\overline{u}}}=\overline{u}\ {%
\mathbf{i}}\,.
\end{equation*}%
Let us note that:
\begin{equation*}
\frac{\partial \mu _{o}(\rho^{\ast } _{b_{x}})}{\partial x}=\frac{\partial
\mu _{o}(\rho^{\ast } _{b_{x}})}{\partial \rho^{\ast } _{b_{x}}}\,\frac{%
\partial \rho^{\ast } _{b_{x}}}{\partial h_{x}^{\ast }}\,\frac{\partial
h_{x}^{\ast }}{\partial x}\equiv \frac{1}{\rho^{\ast } _{b_{x}}}\,\frac{%
\partial P(\rho^{\ast } _{b_{x}})}{\partial \rho^{\ast } _{x}}\,\frac{%
\partial \rho^{\ast } _{b_{x}}}{\partial h_{x}^{\ast }}\,\frac{\partial
h_{x}^{\ast }}{\partial x}.
\end{equation*}%
Due to the fact the vapor
bulk pressure $P^{\ast }_{v_{b_{x}}}$ is   constant along  the xylem tube,   by using relation $\Pi (h_{x}^{\ast
})=P^{\ast }_{v_{b_{x}}}-P^{\ast }_{b_{x}}$, we get along the flow motion:
\begin{equation*}
\frac{\partial \mu _{o}(\rho^{\ast } _{b_{x}})}{\partial x}=-\frac{1}{\rho^{\ast }
_{b_{x}}}\ \frac{\partial \Pi (h_{x}^{\ast })}{\partial h_{x}^{\ast }%
}\ \frac{\partial h_{x}^{\ast }}{\partial x}
\end{equation*}%
and consequently,
\begin{equation}
\chi^{\ast } _{b_{x}}{\mathbf{\overline{u}}}=h_{x}^{\ast }\left( \frac{%
h_{x}^{\ast }}{3}+L_{s}\right) \left[ \,\text{grad}\ \Pi (h_{x}^{\ast })+g\,\textbf{i}%
\right] ,  \label{variation potentiel chimique}
\end{equation}%
where $\chi^{\ast } _{b_{x}}=\rho^{\ast } _{b_{x}}\nu\, $ is
the liquid kinetic viscosity. Consequently, the mean liquid velocity is
driven by the variation of the disjoining pressure along the solid wall and
the film thickness. Equation (\ref{variation potentiel chimique}) differs from the classical film hydrodynamic one. Indeed, for a classical
thin liquid film, the Darcy law is ${\mathbf{\overline{u}}}=-K(h)\,\mathrm{%
grad}\,p$, \ where $p$ is the liquid pressure and $K(h)$ is the permeability
coefficient. In Eq. (\ref{variation potentiel chimique}), the sign is
opposite and the liquid pressure is replaced by the disjoining pressure.  We note that $ \chi^{\ast } _{b_{x}}\simeq \chi$\,, where $\chi$ is the liquid kinetic viscosity in the liquid bulk at phase equilibrium.
Moreover, when $h_{x}^{\ast }/L_{s}\ll 1$,
\begin{equation*}
\chi\,{\mathbf{\overline{u}}}=h_{x}^{\ast }\ L_{s}\left[ \,%
\text{grad}\ \Pi (h_{x}^{\ast })+g\,\textbf{i}\right] ,
\end{equation*}%
which is strongly different from the case  $L_{s}=0$ corresponding to the adherence condition: \begin{equation*}
\chi\,{\mathbf{\overline{u}}}=\frac{h_{x}^{\ast 2}}{3}\left[
\,\text{grad}\ \Pi (h_{x}^{\ast })+g\,\textbf{i}\right] .
\end{equation*}%
The mass equation averaged over the liquid depth is:
\begin{equation*}
\frac{\partial }{\partial t}\left( \int_{0}^{h_{x}^{\ast }}\rho \,dz\right) +%
\mathrm{{div}\left( \int_{0}^{h_{x}^{\ast }}\rho \,\mathbf{u}\,dz\right) =0.}
\end{equation*}%
Since the variation of density is small in the liquid nanolayer, the
equation for the free surface is:
\begin{equation}
\frac{dh_{x}^{\ast }}{dt}+h_{x}^{\ast }\ \mathrm{div}{\ \mathbf{\overline{u}}%
}=0.  \label{h-equation}
\end{equation}%
By replacing (\ref{variation potentiel chimique}) into (\ref{h-equation}) we
finally get:
\begin{equation}
\frac{\partial h_{x}^{\ast }}{\partial t}+\frac{1}{\chi}\  %
\mathrm{div}\left\{ h_{x}^{\ast 2}\left( \frac{h_{x}^{\ast }}{3\,}%
+L_{s}\right)  \Big[\,\mathrm{{grad}\,\Pi (h_{x}^{\ast })+g\,\textbf{i}\Big]}\right\} =0.
\label{h-evolution equation}
\end{equation}%
It is easy to verify that Eq. (\ref{h-evolution equation}) is a non-linear
parabolic equation.\\ If ${ \displaystyle \partial \Pi (h_{x}^{\ast })/
\partial h_{x}^{\ast }<0}$ the flow is stable. This result is in accordance
with the static criterium of stability  for thin liquid layers
\cite{Derjaguin}. \newline
When $L_{s}\neq 0$ we notice the flow is multiplied by the factor $%
1+3L_{s}/h_{x}^{\ast }$. For example, when $h_{x}^{\ast }=3\,nm$ and $%
L_{s}=100\,nm$ which is a Navier length of small magnitude with respect with
experiments, the multiplier factor is $10^2$; when $L_{s}$ is $7\,\mu m$ as
considered in \cite{Tabeling}, the multiplier factor is $10^{4}$, which seems possible in nanotube observations \cite{Mattia}. \newline
Equation (\ref{variation potentiel chimique}) yields the flow rate per unit
of length of xylem tubes. We may remark that Eq. (\ref{variation potentiel
chimique}) is mainly valid at the top of highest trees where the xylem tube
network is strongly ramified.  \newline
A main difference between   Poiseuille flow and   motion in a thin film is
the versatility of the liquid layer flow with respect to Poiseuille'. An
hydraulic Poiseuille flow is \emph{very rigid} due to the liquid
incompressibility, the pressure effects are fully propagated in all the
tube. For a thin layer flow, the flow rate can increase or decrease due to
the spatial derivative of $h_{x}^{\ast }$ and strongly depends on the  locally defined
disjoining pressure. Trees can adapt the disjoining
pressure effects by opening or closing the stomatic cells with the object of
changing the evaporation in its leaves so that the bulk pressure in
micropores can be negative and the transport of water can be differently
dispatched in the tree parts; this seems an important aspect of the model. }}

\section{Discussion and conclusions}

Our model is essentially different from the cohesion-tension theory; it allows a new
explanation of biofluidics by  using methods   of the Russian school of
Derjaguin   and  using   non-adherence conditions for nanofluid flows
at xylem walls. It precisely gives  the possible height of the tallest
trees. In the very thin layers, we obtain rates of flows  strongly larger than those obtained with traditional Navier-Stokes models.
The explanation of irrigation of the leaves in the tallest
trees supplementary   justifies our model of biomimetism. \newline
The motor of the sap motion is induced by  the transpiration across
micropores located in tree leaves \cite{Zimm}. It is natural to conjecture
that the diameters of xylem  tubes must be the result of a competition
between evaporation in tubes which reduces the flow of sap  and the flux of
transpiration in micropores inducing the motion strength.\newline
We notice that the negative pressure only appears in  the liquid mother bulk.    The microlimbs,  micropores and
stomates have a diameter a little  smaller than the bubble size considered in the
introduction; consequently, they can be filled without any  cavitation by the liquid mother bulk at a
suitable negative pressure associated with the height of the tallest trees.  \newline
It is interesting to note that if we switch the microtube surfaces to wedge
geometry  or to corrugated surface,  it is much easier to obtain the
complete wetting requirement;  thus,  plants can avoid having very high
energy surfaces. Nonetheless, they are still   internally wet  if crude sap flows pass through
wedge shaped corrugated pores; this fact also answers to questions in \cite{Zimm}. The wedge does not have to be perfect on the
nanometer scale to significantly enhance the amount of liquid flowing at modest pressures corresponding to nanosized planar films. It is
bound to improve on the calculation because it enhances the surface to
volume ratio. In such a case, we remark that the wall boundary can always be
considered as a plane surface endowed with an average surface energy as in
Wenzel's formula \cite{Wenzel}.\\ Finally, it will be interesting  to confirm our theoretical predictions  with additional experimental data.

\end{document}